# OnBodyAI – Why the Next AI Needs a Body

MICHAEL BEIGL, Karlsruhe Institute of Technology (KIT), Germany

**The next AI will not win by knowing more about the world. It will win because it knows you: because it lives on your body, works for you and answers to you.** That is what the term OnBodyAI stands for. It describes what happens when large language models are given a body, and what research and industry need to make that happen. This article is based on my talk in the Industrial Panel at UbiComp/ISWC 2026 in Shanghai (October 13, 2026) and follows its ten key statements.

## 1 AI HAS A BRAIN, BUT NO BODY

"LLMs know everything about the world – and not much about you, right now." That is the starting point. The wearable and ubiquitous computing community has long been working on connecting wearables to language models: Health-LLM (Kim et al., 2024), for example, showed that LLMs can predict health outcomes from wearable data. Google's PH-LLM coaches sleep and fitness from wearable data, as reported in *Nature Medicine* (Khasentino et al., 2025), SensorLM (Zhang et al., 2025) learns a language for sensor signals, and Apple trained a foundation model on behavioral data from the watch in 2025 (Erturk et al., 2025). Sensor data in, advice out.

And yet everyday reality looks like this: if you ask a state-of-the-art assistant whether to push through tonight or take a break, you get a polite but useless counter-question. How did you sleep? Is your heart racing? How stressed are you? The smartest model in the world has to ask me how I feel, while a few centimeters from my eardrum a sensor already knows my heart rate, my sleep, my jaw clenching and the tension in my voice. **Today's AI has a brain, but no body.**

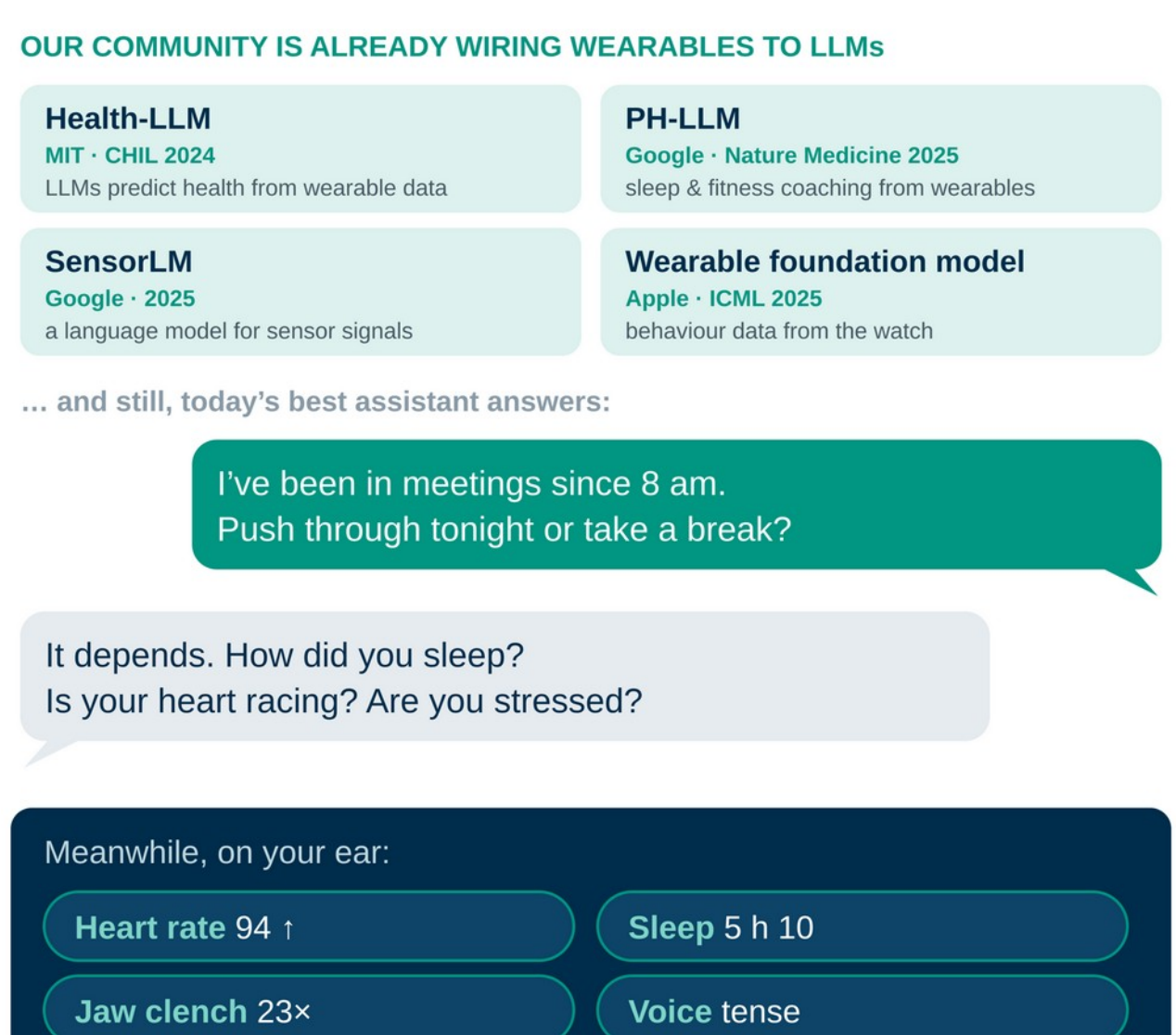

Fig. 1. The assistant asks, while the ear already knows the answer. Own illustration (Beigl/TECO).

Author's address: Michael Beigl, Karlsruhe Institute of Technology (KIT), Karlsruhe, Germany.

For industry the consequence is this: personalization based on physical context is the next differentiator of AI, and body data is the gap that so far has been only weakly occupied.

## 2 THE OTHER HALF: WE HAD THE WEARABLE, WE LACKED THE BRAIN

"In 1999 we could already build context-aware things. What we lacked was the brain." The MediaCup (Gellersen et al., 1999), presented at HUC '99, the predecessor of today's UbiComp, carried sensors, radio and battery in the base of a coffee cup. It knew when the coffee was hot and when a meeting was going on. But everything relied on hand-written rules. It could not learn anything new.

In the 25 years since, our community has learned to read the body like no one else: which signals exist, where to measure them and how to build devices that people actually wear. Today OpenEarable 2.0 carries seven sensors on the ear, and the interpretation is learned and runs on the device. What was missing is the brain, and AI now provides it. **AI has a brain but no body; we have the body. Together, that is OnBodyAI.**

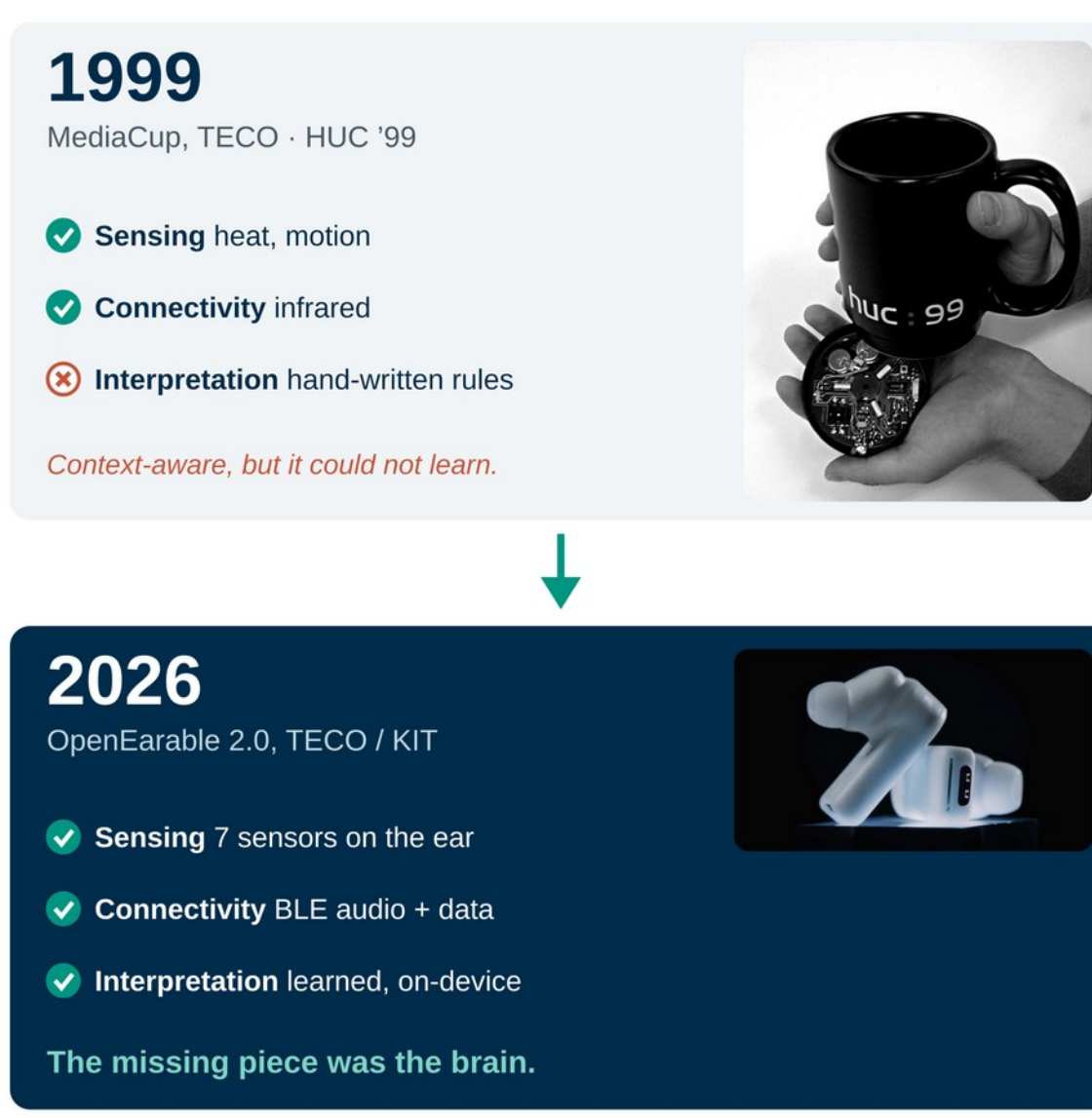

Fig. 2. From the MediaCup (HUC '99) to OpenEarable 2.0. Own illustration and devices of TECO/KIT.

Industry perspective: 25-year-old ideas are finally shippable. For on-body AI, now is the product window.

## 3 GREAT POTENTIAL IS NOT YET A SUCCESSFUL PRODUCT

AI may be the biggest opportunity wearables have ever had. But a cool technology with some idea built on top has never been enough to make a wearable a hit. The wearable graveyard is full of brilliant ideas:

- **Google Glass** (2013–15) bet on miniaturization: a display and a camera on your face. It asked people to look different and made everyone around them uneasy. The Explorer program ended in 2015.

- **Microsoft Band** (2014–16) bet on health sensing, but was stiff and uncomfortable to live with. It was discontinued in 2016.
- **Humane AI Pin** (2024–25) bet on AI itself: talk to your lapel, read a projection on your palm. In February 2025, less than a year after launch, the product was shut down when the company was sold to HP.

Each of these rode a real technology wave, and each asked people to change. **The winners instead upgrade what people already wear:** earbuds that now measure heart rate, rings for sleep, watches for health. Same object, same habit, new intelligence.

From an industry perspective, intelligence therefore belongs in form factors people already wear – no new device, no new habit.

## 4 THE EAR IS THE NEW WRIST

One of these objects is especially interesting: the ear. IDC (2024) counted about 302 million hearables out of roughly 504 million wearables shipped in 2023, that is **about 60 percent**. In September 2025 Apple added heart-rate sensing and live translation to the AirPods Pro 3. An installed base of around 300 million earbuds a year is waiting to become a sensing platform.

Research is following this trend. In our open database EarXplore, publications on ear-based gestures grow from year to year.

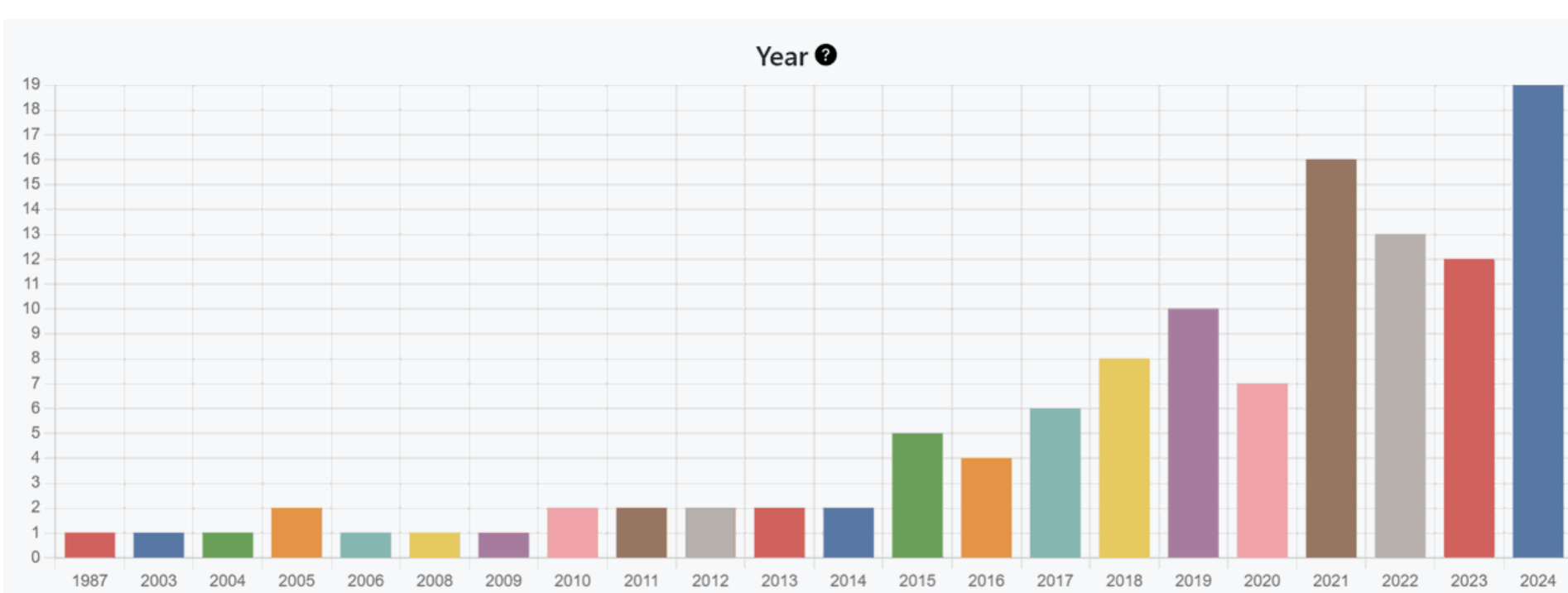

Fig. 3. Papers on ear-based gestures per year. Source: EarXplore (Hummel et al., 2025), own analysis.

Why the ear? It is close to vessels, jaw, eyes and brain, it sits stably, and people wear earbuds for hours anyway. At the ear you are inside and outside the body at the same time, like at few other places on the body. Vessels provide pulse, oxygen saturation and blood pressure; bones carry voice, chewing and steps; muscles – jaw, face, even the tensor tympani – reveal tension; and electrical signals (EOG, EEG) can be derived from eyes and brain. Our systematic review found more than 30 phenomena that can be sensed at the ear (Röddiger et al., 2022). **The body is the richest sensor AI has never had, and the ear is the ideal place for measuring such body signals.**

From an industry perspective, clinically capable signals from consumer earbuds open up the health market – if they hold up all day.

## 5 BEING ON THE BODY IS NOT THE SAME AS UNDERSTANDING THE BODY

A reality check with three numbers from the literature:

- **Sleep stages:** Six commercial wrist wearables reach a Cohen's kappa of only 0.21 to 0.53 against polysomnography. That is not random guessing, but only fair to moderate agreement, far from "substantial" and, above all, below what medical diagnoses would require (Schyvens et al., 2025).
- **Energy expenditure:** A systematic review of 65 studies found a mean absolute percentage error above 30 percent for every wearable brand tested (Germini et al., 2022).
- **Blood glucose from PPG,** an example from our own research: models that look good on randomly mixed data collapse to $R^2 \approx 0$ on people they have not seen, no better than predicting the mean. Yet more than 90 percent of the estimates fall into the "clinically acceptable" Clarke zones A and B, even for the plain mean predictor (Ramesh et al., 2026). **Our metrics hide the failure.**

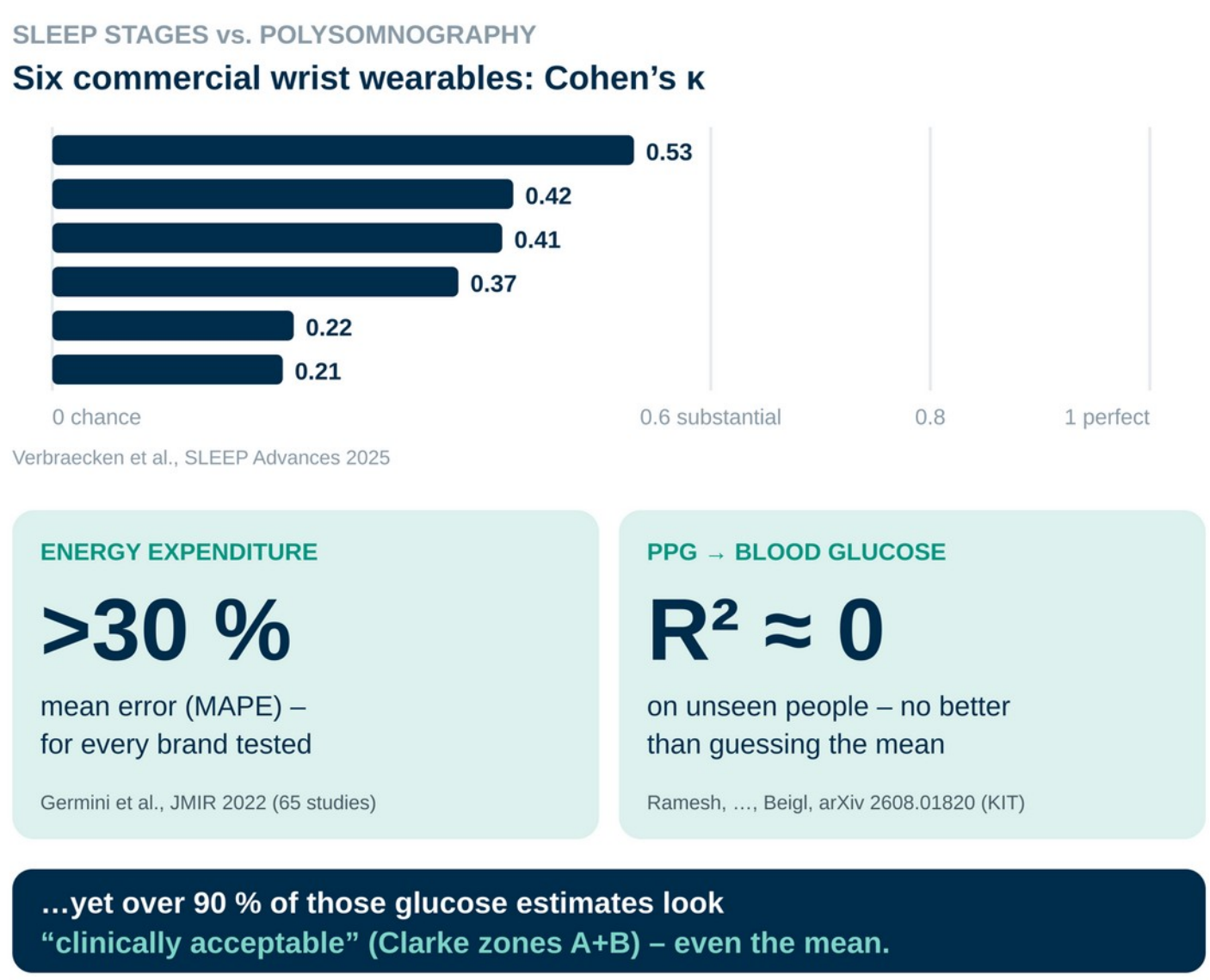

Fig. 4. Where body sensing reaches its limits today. Own illustration based on the studies cited.

AI is often seen as the solution to every problem, but there are limits. **A (bigger) model cannot reconstruct information the sensor never captured.** Contact, motion and signal quality limit what any model can know. Coupling and signal quality are therefore AI problems, not implementation details, and every model must be validated on unseen people, situations, recording environments and so on, not just on random splits. Only then will generalizable "foundation" models emerge.

## 6 CLOSING THE LOOP: AI GIVES YOU FREEDOM

Because sensing is imperfect, acting on the body must happen on the user's terms. When it does, something remarkable happens: **the system does what I want before I have to think of it – and relieves me of cognitive effort.** The loop runs on my body, under my rules, without outside influence, and takes load off me. Two closed loops from our lab show what this looks like:

- **Calmables:** A smart ring measures heart rate, and infrared earables gently warm the ear when the heart rate indicates a need to relax. Sense on the finger, act at the ear.

- **EarStreAM:** OpenEarable 2.0 measures heart rate and heart-rate variability and detects rising stress. An LLM then generates a meditation that adapts in real time to how the body responds.

The language model no longer has to ask me how I feel. It reads it from my body and acts, under my rules. Trust in the system is, however, decisive for whether users hand over control.

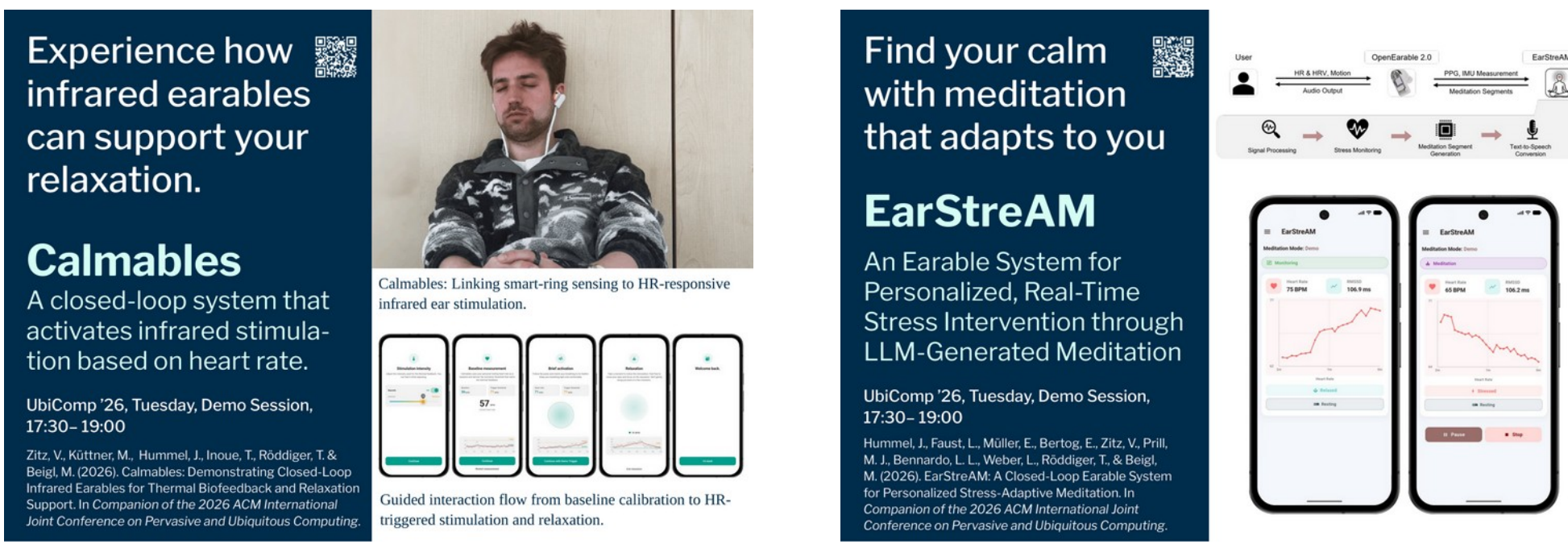

Fig. 5. Two closed loops from our lab. Posters of the UbiComp/ISWC 2026 demo session (Calmables: Zitz et al., 2026; EarStreAM: Hummel et al., 2026). Own illustrations.

The closed loop is currently the subject of both products and research. **Elemind**, an MIT spin-off, has sold an EEG headband since 2024 that measures brain waves in real time and plays short sounds via bone conduction, timed to the phase of your own alpha waves; once sleep is detected, it fades out. Apple's **AirPods Pro** became the first FDA-authorized over-the-counter hearing aid software in September 2024: a hearing test in the earbud tunes the amplification to your own ears. Hundreds of millions of earbuds can thus act medically on the body. And research is moving on: our **Heatables** warm the ear canal with infrared LEDs, and participants felt about 1.5 °C warmer (Zitz et al., 2025). Thermal comfort could be the next feature in an earbud.

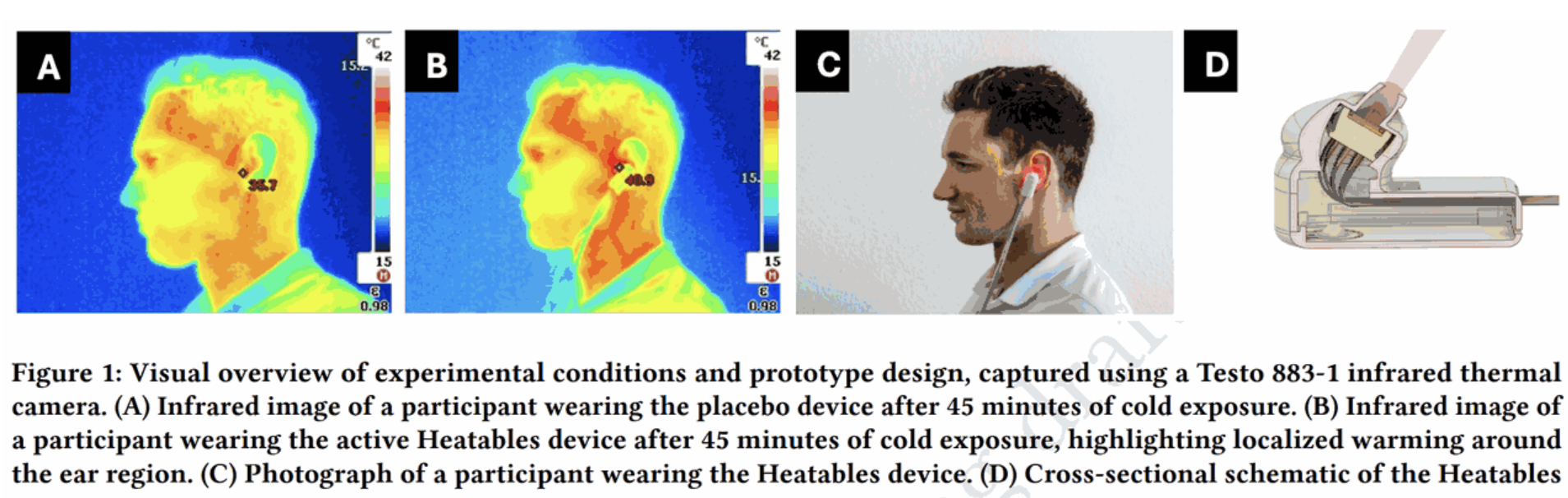

Fig. 6. Heatables, experimental conditions and prototype (Zitz et al., 2025, Fig. 1).

From an industry perspective, this opens up an interesting potential: head and ear become output channels – for sleep, hearing and comfort. It is important to note that users buy outcomes, not sensors, so the channel must deliver lasting value to the user, otherwise a product will not succeed.

## 7 ON THE BODY, MILLIWATTS MATTER MORE THAN MODEL PARAMETERS

A wearable must run all day on a battery the size of a fingernail. A data-center GPU draws around 700 watts, a smartphone SoC a few watts. A whole earbud – sensors, radio, audio and compute together – has on average about 30 milliwatts if it is to last a working day. That is four orders of magnitude of difference.

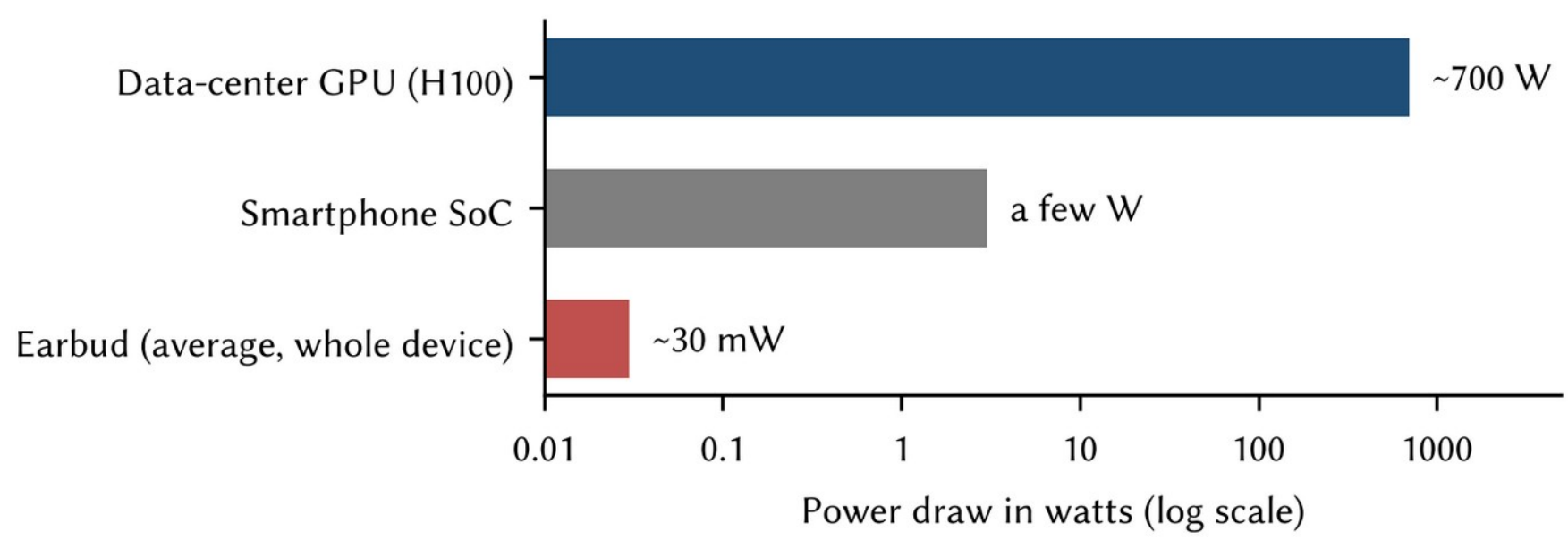

Fig. 7. Power budget comparison (own illustration, Beigl/TECO). GPU: NVIDIA H100 SXM TDP; earbud: ~50 mAh battery over ~6 h of use.

Every additional milliwatt costs heat that must be dissipated on or near the skin, and volume on the body, for example in the ear. Simply shrinking AI models, for example by quantization and pruning, is not optimal, because the properties of the embedded processors used are not sufficiently taken into account. This concerns both the runtime of the processor operations used and their energy consumption, since the model was designed for a different processor. Instead, it is necessary to have the AI model designed (automatically) specifically for the target processor. Our MicroNAS (King et al., 2025) approach is a first step in this direction. From an industry perspective this step is critical: battery life sells wearables, because users are very demanding in this area.

## 8 AI EATS SOFTWARE. WEARABLES ARE STILL BUILT BY HAND

App code, backends and even ML pipelines are increasingly written with AI assistance. But a wearable is not an app. **Building one is a chain of hand work, interrupted by reality at every step.** It often starts with soldering fine-pitch parts under the microscope, or at the latest with testing factory-made prototypes. Often the circuit initially does not behave as a simulation would suggest. Or a part you designed in is no longer available, the substitute is a fraction of a millimeter larger, and suddenly two components are in each other's way. And so on. On top come further real-world influences, such as noise on a line where there should be none. One more hand-soldered fix wire, and again.

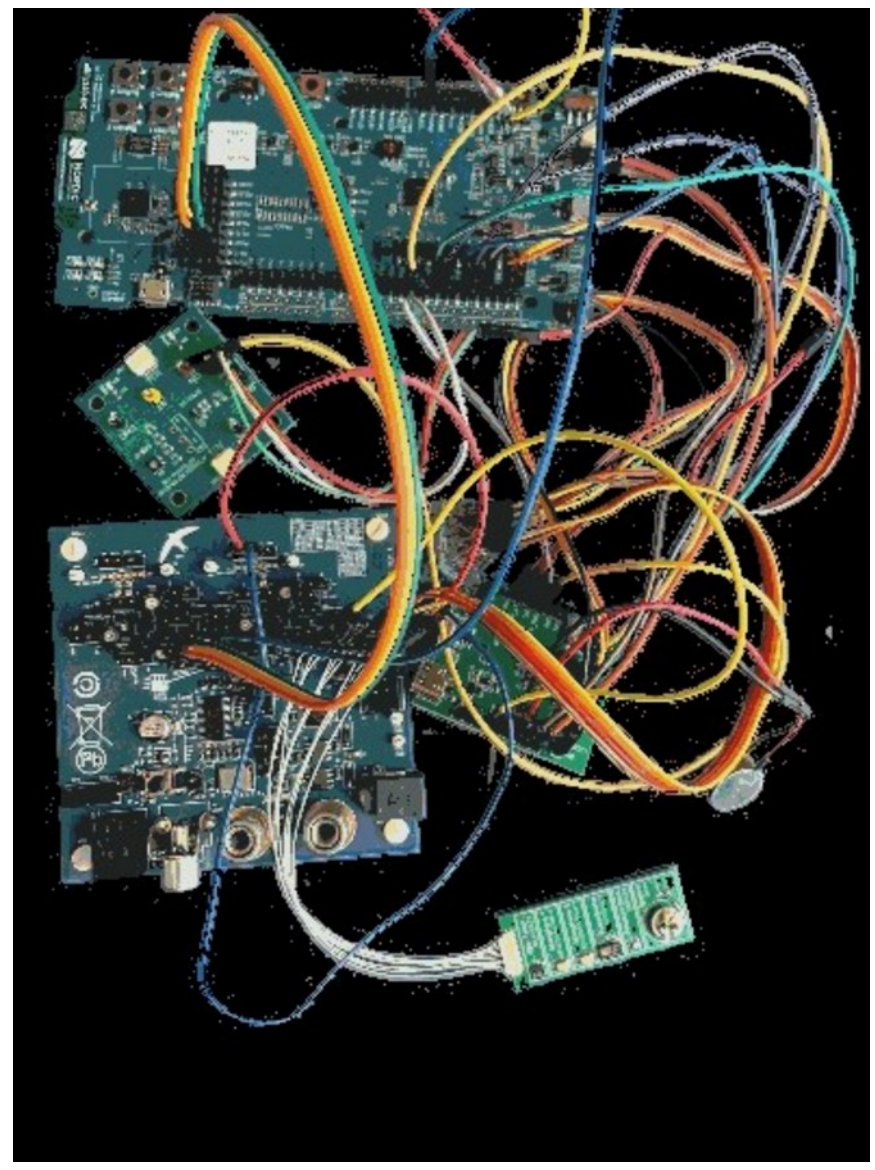

Fig. 8. A typical intermediate prototype state in the lab (TECO/KIT).

No computer does this chain, and (not yet) any robot can support many of the important steps. Most of it is human labor, and it is scarce: embedded engineers are among the hardest to find and most expensive engineers (as David Cuartielles of Arduino put it in his keynote at the OpenWearables workshop 2025 at UbiComp/ISWC). Wearables profit enormously from AI, but building them remains the bottleneck. Our answer is to automate at least part of the embedded stack. **edge-ml.org** takes you in the browser from recording to a model running on the wearable, without writing firmware. The biggest lever of all: do not start from scratch, share the hardware. For industry, automating here is decisive for cost and time-to-market – and open hardware accelerates it further.

## 9 SIGNALS BECOME MEANING – BUT ONLY IF THE HARDWARE PROVIDES THE BASIS

Body signals are weak, noisy and personal: motion artifacts, individual anatomy, noise. This is exactly where learned models are strong. Three examples from our lab:

- **EarRumble:** Between 36 and 50 percent of people can voluntarily contract the tensor tympani, a muscle in the middle ear. It makes a dull rumble that an in-ear pressure sensor picks up. Our classifier reaches 95 percent. Users called it “almost telepathic” (Röddiger et al., 2021).
- **earEOG:** Electrodes around the ear track horizontal eye movement in a normal headphone form factor. Vertical movement does not work yet, and we say so (King et al., 2025).
- **OpenEarable ExG:** Open hardware that captures EEG, EMG and EOG from the ear (Lepold et al., 2024).

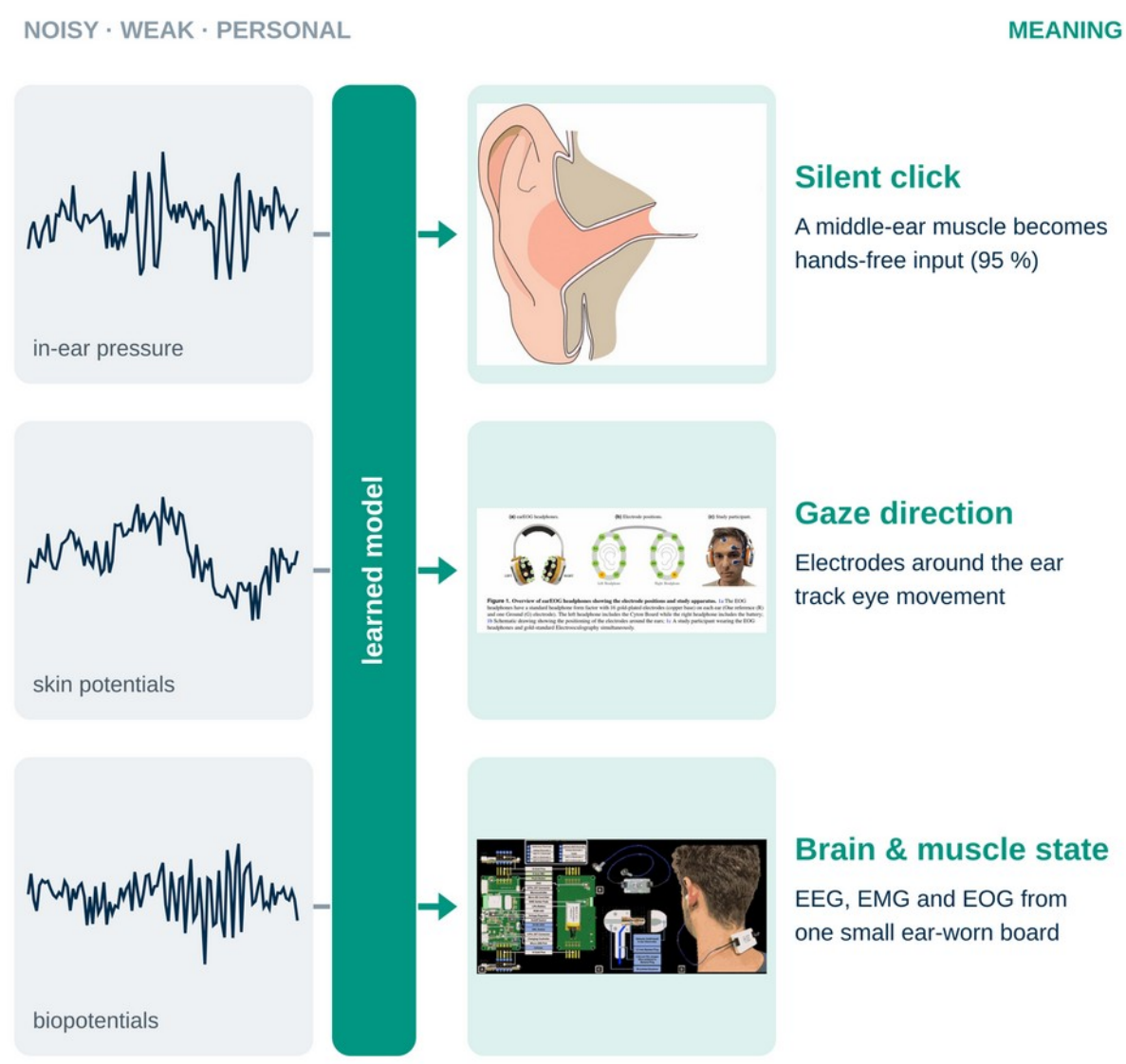

Fig. 9. From signal to meaning. Own illustration based on EarRumble, earEOG and OpenEarable ExG.

But the model can only learn what it gets. A closed earbud gives you steps and heart rate. None of the three examples would have been possible on a closed device. **Hardware quality is the door opener.** For industry this creates a balancing act: raw sensor access is a product decision, and closed interfaces limit what AI can learn. That shuts out competition. In such a dynamic field, however, it also limits the impact of the system.

## 10 CLOSED SYSTEMS GO TO ONE MARKET, OPEN SYSTEMS GO EVERYWHERE

Closed systems reach the market, and that is fine: one vendor, one well-defined product, this is how industry makes its money. Open systems, by contrast, go everywhere: into labs we have never heard of, into clinics, into startups, into applications the vendor never imagined, and into many products. And they create trust, which matters to users (see above). A device that sits on your body all day and listens to your heart and your voice must be trustworthy. **Trust needs openness:** open hardware and software make results reproducible across labs, make privacy claims auditable and speed up product development. Hardware is going the way software went – RISC-V, OpenTitan, TinyTapeout.

Our OpenEarable family is an example: OpenEarable 2.0 is a fully open-source AI platform for ear-based sensing (hardware and software under the MIT license; Röddiger et al., 2025), OpenEarable ExG provides EEG, EMG and EOG from the ear, and OpenEarable BCI extends the platform to brain-computer interfaces.

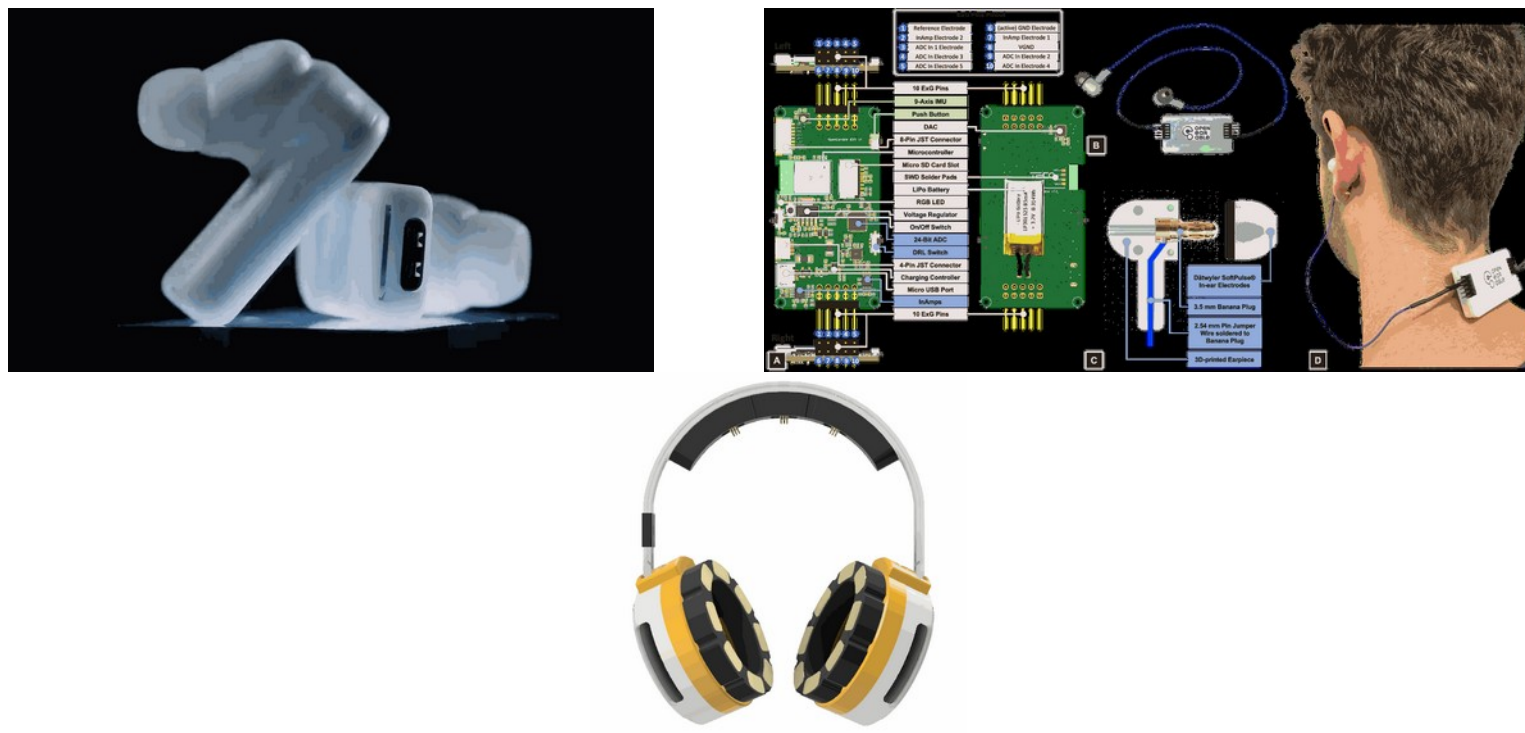

Fig. 10. The OpenEarable family from KIT: 2.0, ExG and BCI. Own devices and illustrations.

Openness is no longer a niche. Across the community there are good examples of devices: the τ-Ring from Tsinghua University (Tang et al., 2025), BioGAP-Ultra from ETH Zürich (Frey et al., 2025) with headband, chest band and EMG sleeve (all design files open), OpenBCI for EEG, and, on the product side, Omi and the PineTime smartwatch as open products. Since 2024 the field has a home in the OpenWearables workshop at UbiComp/ISWC, which we co-organize with Tsinghua, Siegen, UC Davis and DFKI.

### 10.1 Open platforms are the on-ramp to real markets

The idea that openness is the opposite of products is a misconception. Three examples that are not ours:

- **Loop:** People with type 1 diabetes and their families built an open-source closed loop that reads the glucose sensor and doses insulin. In January 2023 it was FDA-cleared as *Tidepool Loop*: from a GitHub repository to a regulated medical product.
- **Nightscout:** Open-source CGM in the cloud. Parents watch their child's glucose live on phone or watch; the community has more than 39,000 members. Their motto: #WeAreNotWaiting.
- **Tympan:** An OSHWA-certified open hearing-aid platform, with schematics and code on GitHub, bringing new hearing algorithms from the lab to listeners.

From an industry perspective, open reference platforms cut time and cost to prototype. As we can see, patients – partly also through the "empowerment" by AI – no longer wait: an open-source system can become a regulated product. Industry has to adapt.

## 11 CONCLUSION: GIVING AI A BODY, AND GIVING PEOPLE THE CONTROLS

The next AI will not win because it knows more about the world, but because it knows you. In an optimal setting:

1. **It "lives" on you.** The body is the context AI is missing, and the ear is where people already wear it.
2. **It works for you.** The closed loop, under your own rules, gives you freedom instead of another app to look at.
3. **It answers to you.** It is open, runs on milliwatts and is honest about its limits: about what it can and cannot sense.

OnBodyAI, as we understand it, is therefore not a product category but a research agenda. Its central claim is that the value of an AI system on the body depends less on the size of its model than on the quality of the coupling between sensing, interpretation and actuation, and on whether this coupling remains under the control of the person wearing it. Making this claim operational requires progress on three fronts at once: sensing hardware that captures physiologically meaningful signals reliably in everyday life, models designed for milliwatt budgets and validated on people they have never seen, and open platforms that make results reproducible and claims auditable. Only when these come together will AI on the body move from collecting data to producing a measurable and lasting benefit for the individual.